\documentclass[preprint]{aastex63}

\shorttitle{Periodic activity of SGR 1806-20}
\shortauthors{Zhang, Tu \& Wang}

\begin{document}

\title{\textbf{Possible} periodic activity in the short bursts of SGR 1806-20: connection to fast radio bursts}

\author[0000-0001-6545-4802]{G. Q. Zhang}
\affiliation{School of Astronomy and Space Science, Nanjing University, Nanjing 210093, China}

\author[0000-0001-6606-4347]{Zuo-Lin Tu}
\affil{School of Astronomy and Space Science, Nanjing University, Nanjing 210093, China}

\author[0000-0003-4157-7714]{F. Y. Wang} \affiliation{School of Astronomy and Space Science, Nanjing
    University, Nanjing 210093, China} \affiliation{Key Laboratory of
    Modern Astronomy and Astrophysics (Nanjing University), Ministry of
    Education, Nanjing 210093, China}

\correspondingauthor{F. Y. Wang}
\email{fayinwang@nju.edu.cn}

\begin{abstract}
Magnetars are highly magnetized neutron stars that are characterized
by recurrent emission of short-duration bursts in soft gamma-rays/hard X-rays.
Recently, FRB 200428 were found to be associated with an X-ray burst from a Galactic magnetar.
Two fast radio bursts (FRBs) show mysterious periodic activity. However, whether magnetar X-ray bursts
are periodic phenomena is unclear. In this paper, we investigate the period of SGR 1806-20 activity.
More than 3000 short bursts observed by different telescopes are collected, including the observation of RXTE, HETE-2, ICE and Konus.
We consider the observation windows and divide the data into two sub-samples to alleviate the effect of unevenly sample.
The epoch folding and Lomb-Scargle methods are used to derive the period of short bursts. We find a possible
period about $ 398.20 \pm 25.45 $ days. While other peaks exist in the periodograms. If the period is real, the connection between short bursts of magnetars and FRBs should be extensively investigated.

\end{abstract}


\section{Introduction} \label{sec:intro}
Soft gamma repeaters (SGRs) are associated with
extremely magnetized neutron stars (named magnetars, \citet{Kouveliotou1998Natur.393..235K,Kaspi2017}).
Magnetars undergo occasional random outbursts
during which time their persistent emission increases
significantly while simultaneously emitting bursts (or intermediate
flares), in the hard X-ray or soft gamma-ray energy regime.
More than 20 SGRs have been discovered and multiple bursts have been
detected from each source. Recently, SGR 1935+2154 reached
its active phase and produced a burst forest. Among these bursts,
there is a very special burst that is associated with FRB 200428
\citep{TheCHIME/FRBCollaboration2020arXiv200510324T,Bochenek2020arXiv200510828B,Lin2020,Li2020arXiv200511071L,Mereghetti2020ApJ...898L..29M}.

Fast radio bursts (FRBs) are millisecond duration radio bursts with high dispersion measures and brightness temperatures
\citep{Lorimer2007Sci...318..777L,Petroff2019A&ARv..27....4P, Cordes2019ARA&A..57..417C,Zhang2020,Xiao2021}. Among the observed FRBs,
repeating FRBs are more interesting. They show multiple bursts, which indicates a non-catastrophe central engine, such as
the flaring activity of magnetars \citep{Lyubarsky2014MNRAS.442L...9L,Kulkarni2014,Katz2016,Murase2016,Metzger2017,Beloborodov2017ApJ...843L..26B,Wang2020}, the cosmic combing
\citep{Zhang2017ApJ...836L..32Z,Zhang2018ApJ...854L..21Z}, the collision between neutron stars and asteroids
\citep{Geng2015ApJ...809...24G,Dai2016ApJ...829...27D}, etc \citep{Platts2019PhR...821....1P}. Moreover, the statistical properties of the repeating bursts are consistent
with those of Galactic magnetar bursts \citep{Wang2017,Wadiasingh2019,Cheng2020}.
Recently, FRB 200428 has been detected to be originated from the Galactic SGR 1935+2154
\citep{TheCHIME/FRBCollaboration2020arXiv200510324T}. This observation supports the model that
FRBs origin from magnetars. The burst time of FRB 200428 is consistent with that of an X-ray burst \citep{Li2020arXiv200511071L,Mereghetti2020ApJ...898L..29M}.

An intriguing property of repeating FRBs is the mysterious periodic activity.
FRB 180916, the second localized repeating FRB, has been found with a period of $16.35 \pm 0.15$ days
\citep{Chime/FrbCollaboration2020Natur.582..351C}. Later, \citet{Rajwade2020MNRAS.495.3551R} found
a possible period 156 days for FRB 121102, which was confirmed by \citet{Cruces2020}. So far, there is no similar period behavior that have been found
in other repeating FRBs. This may be caused by the small number of observed bursts.
Whether all repeating FRBs are periodic is still unknown.

Due to the connection between radio bursts of FRBs and X-ray bursts of magnetars, it is natural to consider whether
X-ray bursts of magnetars have similar periodic behavior.
Although only one radio burst has been observed from SGR 1935+2154,
many short X-ray bursts have been detected from this source. A possible periodic behavior
has been found in SGR 1935+2154 \citep{Grossan2020arXiv200616480G}. The reported period is about 232 days, which is similar to that of FRB 121102.
The spin period of SGR 1935+2154
is about 3.2 s \citep{Israel2016MNRAS.457.3448I}, which is much shorter than 232 days.
Thus, this active cycle must be caused by other processes.

There are some
models to explain the periodic behavior of FRBs. The first one is a binary system containing a magnetar
\citep{Ioka2020ApJ...893L..26I,Lyutikov2020ApJ...893L..39L,Gu2020}. The periods of FRBs origin from
the orbital periods of the binaries. The second model is the free precession of magnetars
\citep{Levin2020ApJ...895L..30L,Zanazzi2020ApJ...892L..15Z}. In this scenario, the strong magnetic field
deforms the magnetar, which induces the free precession with a period from weeks to months. Similar to the free
precession, there are also some works to investigate the force precession. \citet{Sobyanin2020MNRAS.497.1001S} suggested
that the forced precession is natural and can be used to explain the period of FRBs.
The fallback disk and the orbit motion may also induce the force precession \citep{Tong2020arXiv200210265T,Yang2020ApJ...893L..31Y}.
Besides, the magnetar-asteroid impact model also is proposed
to explain the observed periodicity \citep{Dai2020}.

Although a possible periodic behavior has been found in SGR 1935+2154, it is still unclear that whether the periodic
behavior is common in magnetars. If the active cycle is unique for SGR 1935+2154, the origin of this behavior
may be associated with the birth of FRBs, as suggested by some binary models \citep{Ioka2020ApJ...893L..26I}.
While if the period behavior is common for magnetars, the mechanism of this period maybe also valid for FRBs.
SGR 1806-20 is a typical magnetar, which was discovered in 1979. Until to now, thousands of bursts have been detected
from this source \citep{Ulmer1993ApJ...418..395U,Aptekar2001ApJS..137..227A,Nakagawa2007PASJ...59..653N,Prieskorn2012ApJ...755....1P}.
We investigate the periodic behavior of this active source.

This letter is organized as follows. In Section \ref{sec:data}, we compile the observations of SGR 1806-20 which are used to derive the period.
In Section \ref{sec:method},
two methods are used to derive the period. We discuss
the possible relationship between FRB period and SGR period in Section \ref{sec:diss}. Finally,
conclusions are given in Section \ref{sec:conclusion}.

\section{The data sample} \label{sec:data}
After the first detection in 1979, SGR 1806-20 has been observed by many telescopes. Thousands
of bursts have been reported \citep{Ulmer1993ApJ...418..395U,Aptekar2001ApJS..137..227A,Nakagawa2007PASJ...59..653N,Prieskorn2012ApJ...755....1P}.
The spin period of this source is 7.55 s and the spin-down rate is $ 4.95\times 10^{-10} $ s/s \citep{Woods2007ApJ...654..470W}.
Among the observed magnetars, the surface magnetic field of SGR 1806-20 is the strongest, which is about $ 2\times 10^{15} $ G
\footnote{\url{http://www.physics.mcgill.ca/~pulsar/magnetar/main.html}}\citep{Olausen2014ApJS..212....6O}. The strong dipole
field is capable to drive strong bursts. As an example, a giant flare has been detected from this source \citep{Palmer2005Natur.434.1107P}.
No FRBs-like event was detected to associated with this giant flare \citep{Tendulkar2016ApJ...827...59T}.

The observations of SGR 1806-20 from different telescopes are collected.
This source is close to the ecliptic, so it is difficult to observe in December and January. The unevenly sample may induce the
false periodic signal. To alleviate this effect, we also collect the observation windows of different telescopes.
The collected data includes the following four sub-samples.
\begin{itemize}
    \item The observation of Rossi X-ray Timing Explorer (RXTE).
          We use the catalog reported by \citet{Prieskorn2012ApJ...755....1P}, which contains over 3040 bursts from SGR 1806-20. This catalog collects the bursts
              observed from Nov. 1996 to Sep. 2009. The timeline of RXTE is recorded in \textit{XTEMASTER}
              \footnote{\url{https://heasarc.gsfc.nasa.gov/W3Browse/all/xtemaster.html}} and
              \textit{XTESLEW}\footnote{\url{https://heasarc.gsfc.nasa.gov/W3Browse/xte/xteslew.html}} catalog. We collect the observation
          windows from these two catalogs.
    \item The observation of The High Energy Transient Explorer (HETE-2). 50 bursts are recorded in this sub-sample
          \citep{Nakagawa2007PASJ...59..653N}. The observation last for five years, from 2001 to 2005. Among these bursts, 41 bursts are detected in 2004 and 2005.
          HETE-2 is always point in the anti-solar direction. Therefore, the bursts observed by HETE-2 are concentrated on the summer season. The timelines
          of HETE-2 are listed in \textit{HETE2TL}\footnote{\url{https://heasarc.gsfc.nasa.gov/W3Browse/all/hete2tl.html}} catalog. However, this catalog only lists the observation
          time, not the duration of observation. In our calculation, we only consider the most important observation windows.
          We assume that the observation during summer is continuous
          and use the timeline recorded in \textit{HETE2TL} to set the start time and end time of each year.
    \item The observation of Konus-Wind. This sub-sample includes 25 bursts \citep{Aptekar2001ApJS..137..227A}. In this sample,
          only one burst occurred in 1979, the other bursts occurred in 1996-1999. We delete
          	the 1979 burst from this sub-sample because it was observed by another telescope. The entire celestial is monitored by Konus-Wind with a duty cycle of 95\%. In our calculation, we assume that the observation
          of Konus-Wind is continuous\footnote{V. D. Pal'shin private communication}. The earliest burst and the latest burst are taken as the start time and end time of this observation window.
    \item The observation of Internal Cometary Explorer (ICE). It contains 134 bursts from 1979 to 1984 \citep{Ulmer1993ApJ...418..395U}.
          Most of the observed bursts occurred in 1983. This telescope is designed to continuously observe the Sun. It can continuously observe
          any source closed in the ecliptic plane. \citet{Laros1987ApJ...320L.111L} estimated the effective coverage from 1978 Aug. to
          1983 Dec. to be 75\% $\pm$ 25\% and after 1984 Jan. to less than 20\%. This telescope has expired many years. We are unable to
          obtain the detailed observation windows. For simplification, we assume the observation is continuous until Jun. 1984. This
          hypothesis covers the duration with high effective coverage and contains all the bursts. 
\end{itemize}
We divide these sub-samples into two
classes: sample A, the observation of RXTE and sample B, the other three observations.
The sample A contains more bursts and has a clear observation window, which is used to derive the period of SGR 1806-20.
The sample B is used to examine the reliability of the period derived from the sample A.

\section{Methods} \label{sec:method}
Two methods are used to search the period of SGR 1806-20, including the epoch folding and the Lomb-Scargle periodogram.
In our calculation, MJD 43840 is taken as phase 0. This choice is a little arbitrary, but it does not significantly affect our results if the period is real.

\subsection{Epoch Folding}\label{subsec:epoch}
The epoch folding method has been used to derive the period of FRB 180916 \citep{Chime/FrbCollaboration2020Natur.582..351C}.
We try to find the active period of SGR 1806-20 using this method. The burst time of SGR 1806-20 can be folded into different
phases through
\begin{equation}
    \label{eq:fold}
    \psi = \frac{T - T_0}{P} - floor(\frac{T - T_0}{P}),
\end{equation}
where $\psi$ is the folded phase, $T$ is the burst time, $T_0$ is the start point (MJD 43840), $P$ is a given period,
and $floor$ is a function which returns the floor of the input number. The folded phases $\psi$ are grouped
into different phase bins. For sample A, we use 20 phase bins in our calculation. The number of bursts in sample B
is less, so we only use 10 phase bins. The classical Pearson $\chi^2$ is used
to examine the derivation from uniformity. It can be calculated as
\begin{equation}
    \label{eq:chi2}
    \chi^2 = \sum^{20}_i \frac{(N_i - pT_i)^2}{pT_i},
\end{equation}
where $N_i$ is the observed number of bursts in the $i$th bin, $p$ is the average burst rate, and $T_i$ is the
observation time of the $i$th phase bin. The peak of $ \chi^2 $ indicates the period of the source.

We search the period from 50 days to 500 days with the step $\Delta f = 0.1/ T_{span} $ in frequency, where $T_{span}$ is the longest time
between the first observed burst and the last one.
The reduced $\chi^2$ of these two samples are shown in the top panels of Figure \ref{fig:sampleA} and Figure \ref{fig:sampleB},
respectively. The vertical green dashed line indicates the peak of the reduce $\chi^2$. Using sample A, we derive the
period is about 398.20 days. However, there is a peak around 430 days with similar significance. We use the vertical red dashed line
to indicate this peak. This period is caused by the observation window, we will prove it using Lomb-Scargle periodogram in the following section.
The peak of the reduced $\chi^2$ of sample B is 395.86 days, which is consistent with that of sample A. We use the vertical green dashed
line to indicate this peak and use the vertical red dashed line to denote a similar high peak, which is caused by the observation window.
In these two samples, the high reduced $ \chi^2 $ of the peak suggest that this peak is almost impossible to be caused by chance. But it can not rule
out the case that these peaks are caused by the observation windows.


\subsection{Lomb-Scargle periodogram} \label{subsec:LS}
The Lomb-Scargle periodogram can be used to deal with unevenly sampled observations \citep{Lomb1976Ap&SS..39..447L,Scargle1982ApJ...263..835S,VanderPlas2018ApJS..236...16V}.
\citet{Cruces2020} used this method to verify the period of FRB 121102.
We use the \textit{LombScargle} function provided by \textit{astropy}\footnote{\url{https://www.astropy.org/}} to calculate the periodogram
of these two samples. The period is searched from 50 days to 500 days.
The periodograms are shown as blue solid line in the middle panels of Figures \ref{fig:sampleA} and \ref{fig:sampleB}, respectively. The vertical green
dashed lines indicate the periods derived from the epoch folding method. In each figure, the green dashed line is coincided with one
peak of the Lomb-Scargle periodogram. However, there are other peaks in each periodogram. The periodogram of sample A has a maximum peak about
 76.96 days, which is caused by the observation window. The peak of sample B is 395.86 days, but there are some peaks  similar
 significance.

We also check the false alarm probabilities of the peaks, which are $2.11 \times 10^{-84}$ for the peak 398.20 days in sample A and $1.45 \times 10^{-40}$ for the peak 395.86 day in
sample B. This low false alarm probability suggests that these peaks are unlikely to have occurred by chance. However, it may be caused by the observation window,
not the internal period of SGR 1806-20.

\subsection{Simulation}
\begin{figure}
    \centering
    \includegraphics[width=0.6\linewidth]{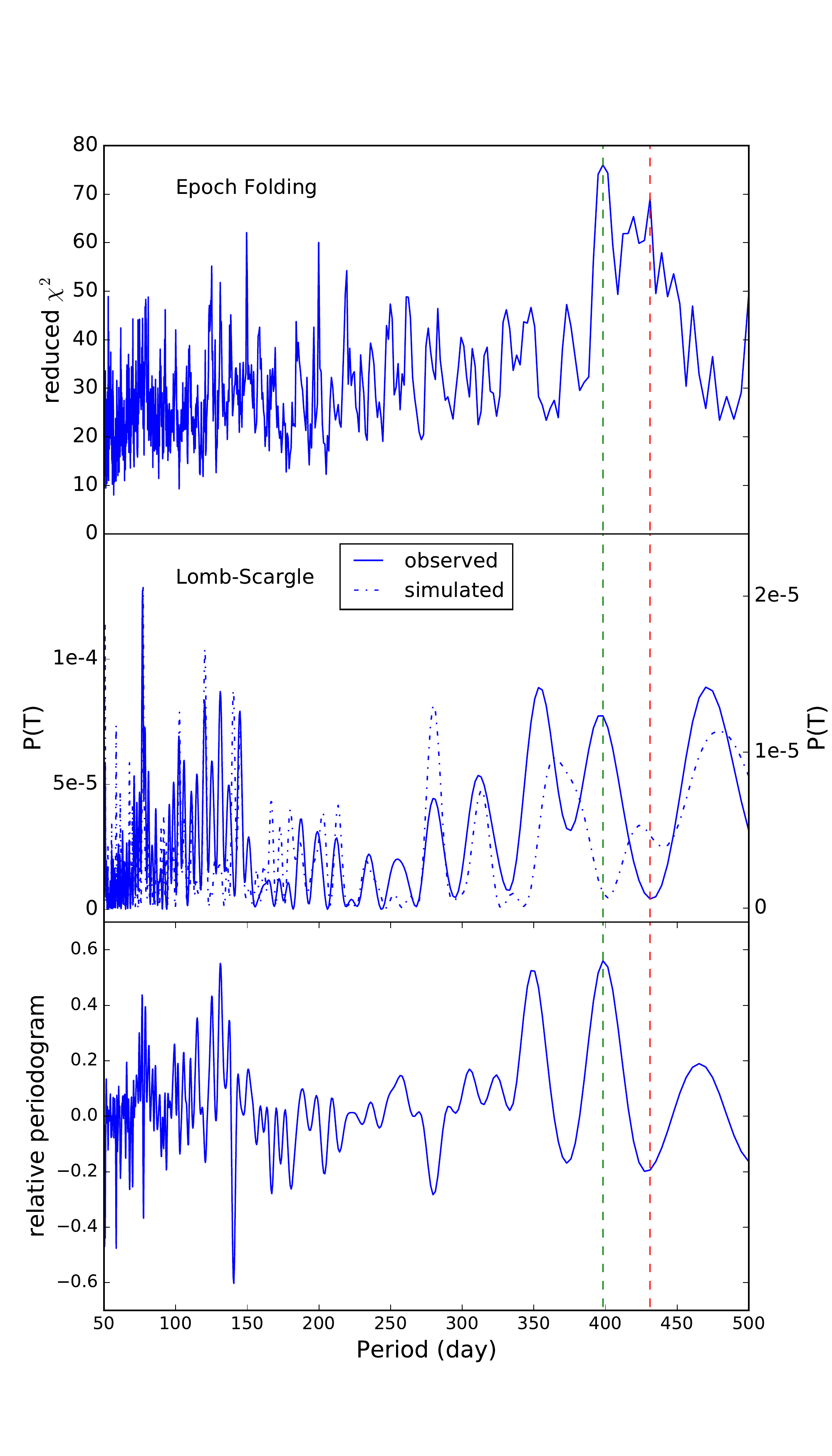}
    \caption{The period of SGR 1806-20 derived from sample A. This sample contains the observation of RXTE, which has
    	over 3040 bursts. The top panel is the result of epoch-folding method. The Lomb-Scargle periodogram
    	of observed data and simulated data are shown in the middle panel with blue solid line and blue dot-dashed line, respectively.
    	The simulated data are uniformly  distributed in
    	the observation windows, so these peaks are caused by the observation windows. We normalized these two periodgrams with
    the maximum value and subtract the simulated periodogram from the observed one. The result is shown in the bottel panel with blue solid line.
		The green dashed line indicates the period derived from epoch-folding, which is 398.20 day. This period is consistent with
	a peak of Lomb-Scargle method. Considering the peak caused by the observation window, the maximum peak is also consistent with this period. The red dashed line
	is another peak in the reduce $ \chi^2 $, which is consistent with a peak of simulated points. Therefore, this peak is caused by the observation window.}
    \label{fig:sampleA}
\end{figure}

\begin{figure}
    \centering
    \includegraphics[width=0.6\linewidth]{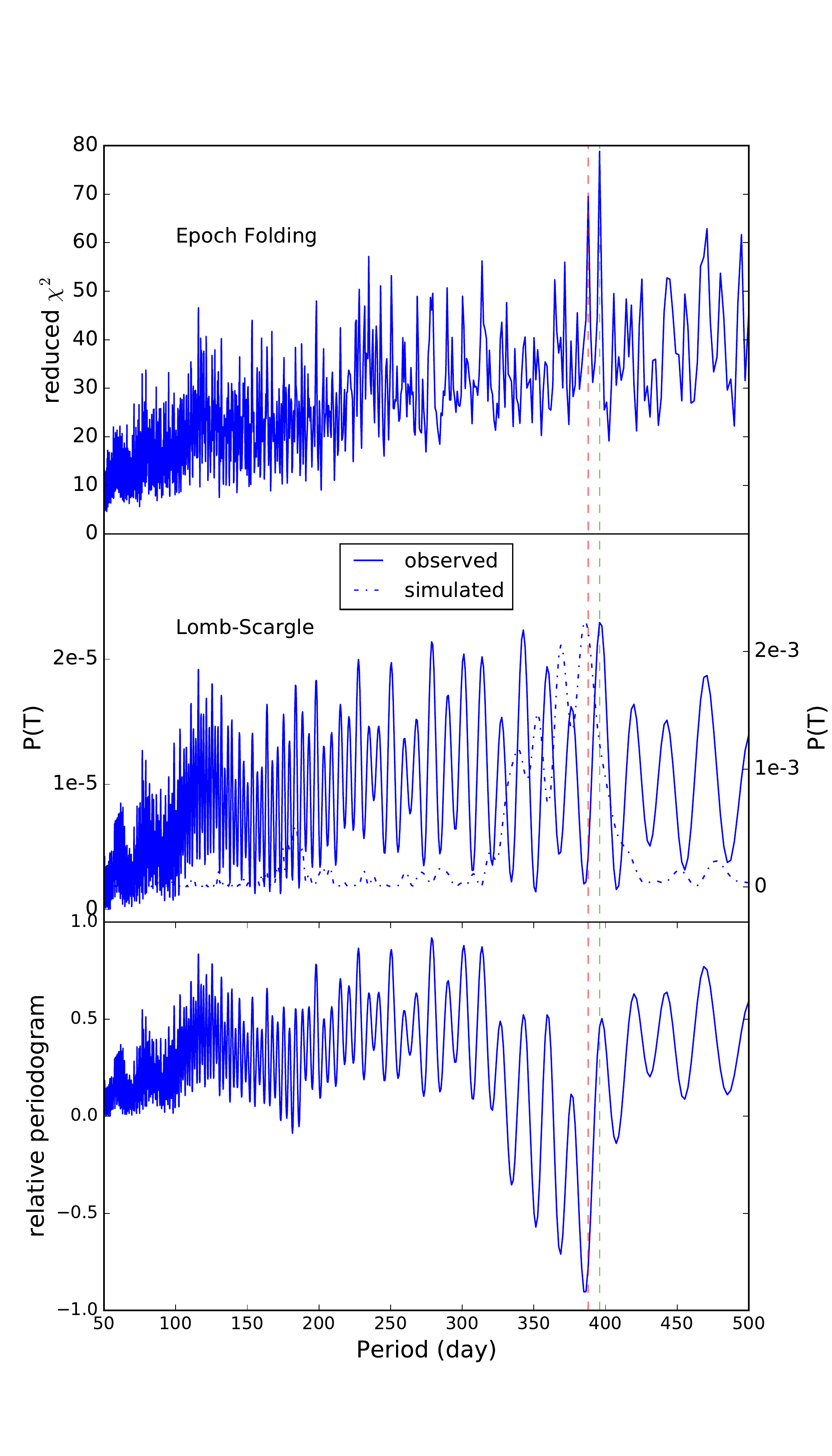}
    \caption{The period of SGR 1806-20 derived from sample B. This sample contains the observations of HETE-2, Konus and ICE.
        It contains 208 bursts. We show the results derived from epoch-folding and Lomb-Scargle periodogram. 
       	We show the reduce $ \chi^2 $ in the top panel. The vertical green line is the derived period 395.86 days.	
        In the middle panel,
        the blue solid line is the observed periodogram and the blue
    dot-dashed line is the simulated periodogram. We normalized these two periodgrams with
    the maximum value and subtract the simulated periodogram from the observed one. The result is shown in the bottel panel.
    The green line indicates the best result of epoch-folding. It is consistent well
        with the period given by the Lomb-Scargle method. The vertical red dashed line is another significant peak of reduce $ \chi^2 $.
    This line is consistent with a peak of simulated periodogram, so it is caused by the observation window.}
    \label{fig:sampleB}
\end{figure}

The observation windows have a strong impact on the period search. To understand its effect, we simulate a
	series of points that are uniformly distributed in the observation windows. The Lomb-Scargle method is used to deal with these simulated
	points. The interval of simulated points is 0.1 days, which is much shorter than the possible period of SGR 1806-20. Therefore,
	the internal period of simulated points would not affect the results. The peaks in the simulated periodogram are caused by the unevenly sample.
	We show the periodogram of simulated points in the middle panels of Figure \ref{fig:sampleA} and \ref{fig:sampleB} with blue dot-dashed
	lines. In Figure \ref{fig:sampleA}, many peaks of the observed data coincide with the peaks of the simulated points. In this figure, the
	vertical red dashed line is the second peak derived from epoch-folding. This line is coincided with a peak of the simulated periodogram,
	which supports that this peak is caused by the observation window. While the green line agrees with a bottom of the simulated periodogram.
	Therefore, this peak is unlikely to be caused by the observation window. We normalize the observed periodogram and the simulated periodogram
	with maximum values and subtract the simulated periodogram from the observed periodogram.
	The result is shown in the bottom panel of Figure \ref{fig:sampleA} as blue solid line. In this figure, the vertical green dashed line is the period
	derived from epoch folding and vertical red dashed line is the second peak of epoch folding results. The peak of this periodogram is 398.20 days, which is
	consistent with the period derived from epoch-folding. We also check the Lomb-Scargle periodogram caused by the observation window of sample B.
 The simulated periodogram is shown in the middle panel of Figure \ref{fig:sampleB} with blue dot-dashed line. This periodogram has several peaks
 near 360 days. Like sample A, we subtract the normalized simulated periodogram from the normalized observed periodogram and show the
 result in the bottom panel of Figure \ref{fig:sampleB}. This periodogram also has a peak about 395 day, but the maximum peak is about 278
 days. This may be caused by two reasons. The first one is inappropriate subtraction. We normalize these two
 periodograms with maximum values and perform the subtraction. It can tell us which peak
 is caused by the observation window, but can't give the significance of this peak. The second reason is the incomplete observation
 window. The detailed observation windows of HETE-2 and ICE are unclear. We assume the observation is continuous in a specific
 window. This assumption imports some uncertainties.

\begin{figure}
    \centering
    \includegraphics[width=\linewidth]{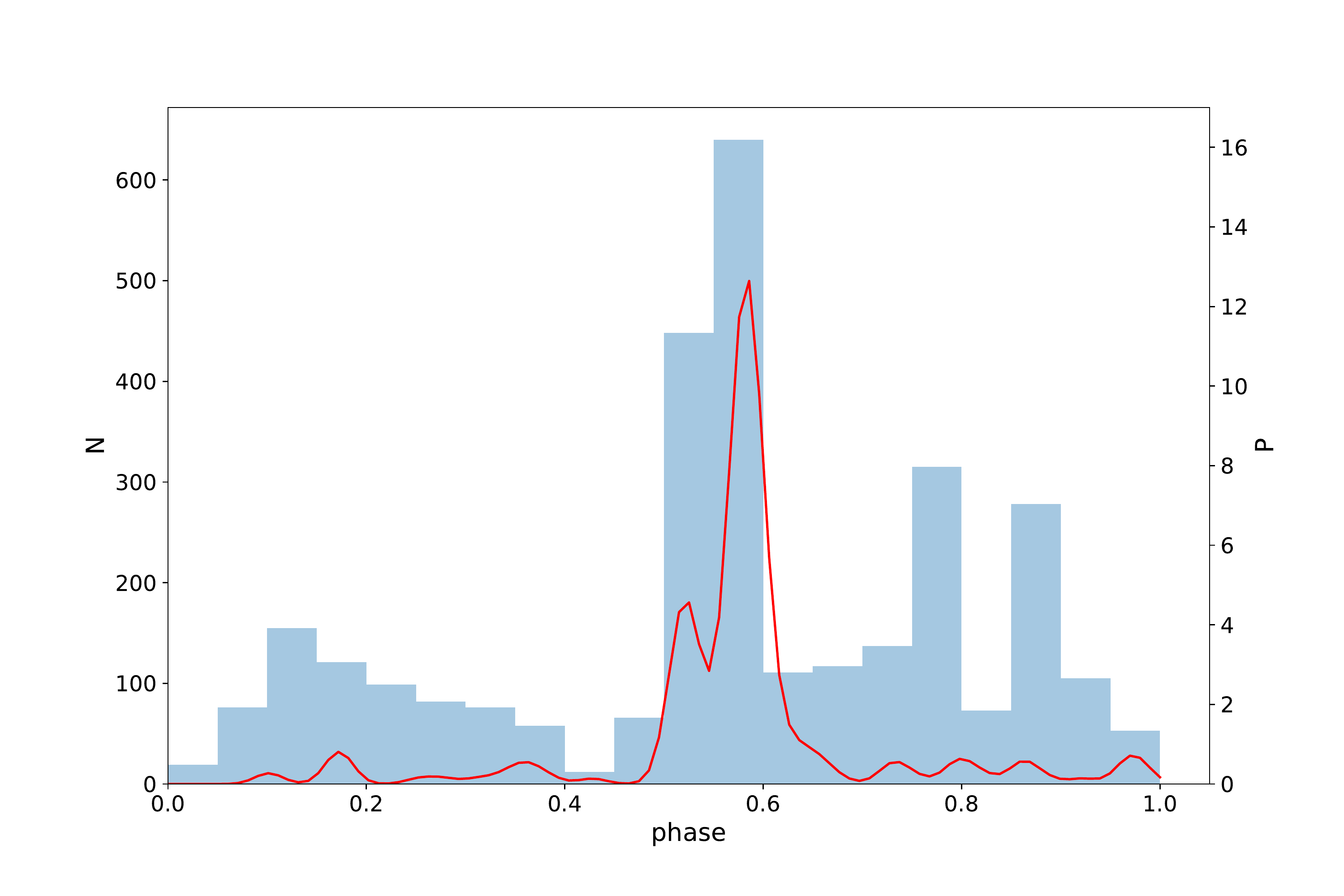}
    \caption{The folded phase of SGR 1806-20. The blue histogram is the distribution of sample A and the
        red line is the kernel density estimate of sample B. These two distributions are consistent with each other
        at the main peak $\psi \simeq 0.58$. In sample A, there are three minor peaks, which are invisible in sample B.}
    \label{fig:foldedphase}
\end{figure}
\begin{figure}[htbp]
    \centering
    \includegraphics[width=\linewidth]{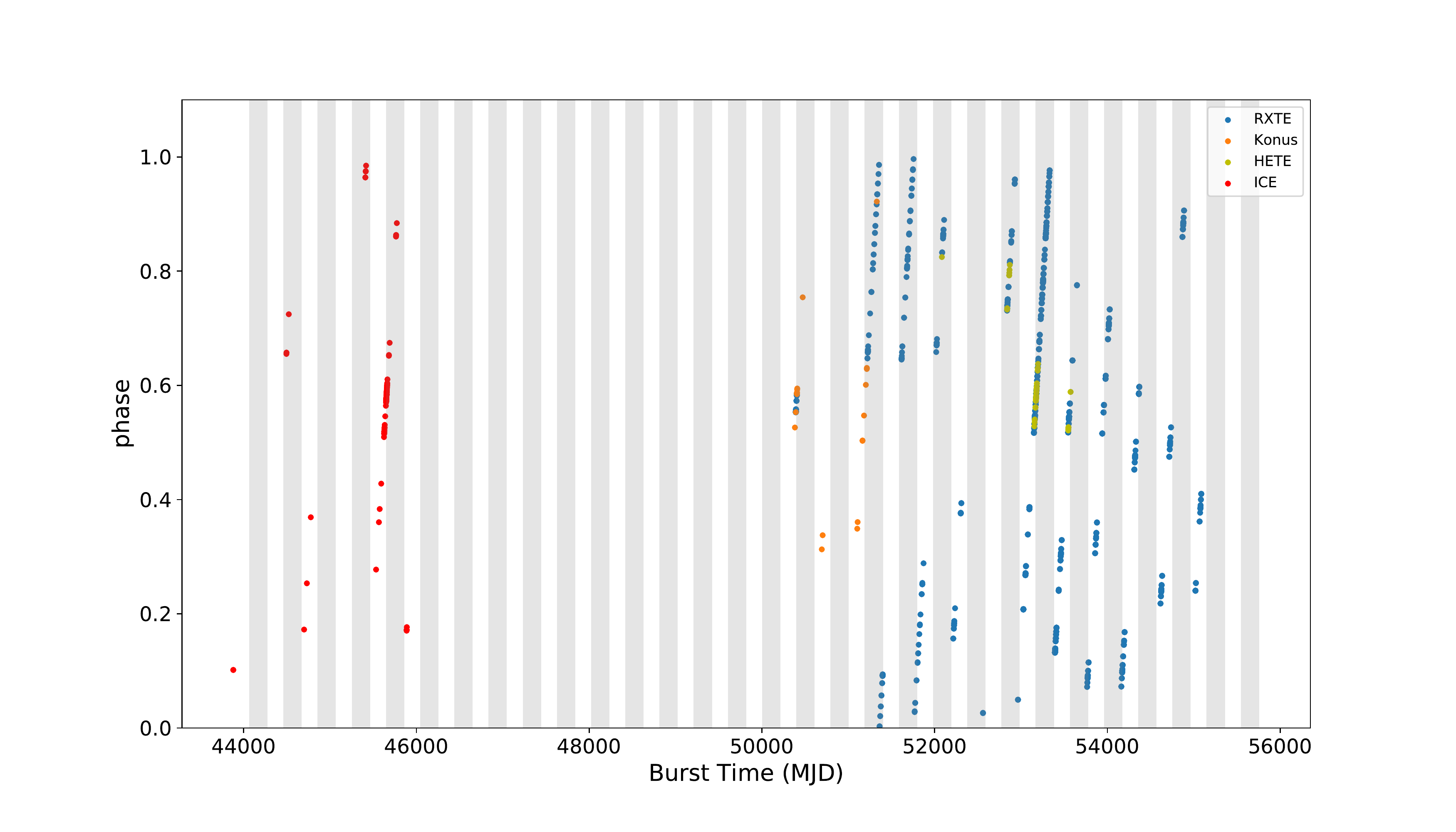}
    \caption{The MJD and phase of SGR 1806-20. The gray
        regions denote the main peak and the later two minor peaks in Figure \ref{fig:foldedphase}. Due to the existence of other minor peaks, there
        are some points outside the gray regions.}
    \label{fig:mjd}
\end{figure}

According to these two methods, the period is about $ 398.20 \pm 25.45 $ days for sample A and $ 395.86 \pm 3.92 $ days for sample B. The error is derived using the method
in \citet{Chime/FrbCollaboration2020Natur.582..351C}. It can be derived as $\sigma_{P} = PW_{active}/T_{span} $, where
$ P $ is the period and  $W_{active}$ is the number of active days. The periods of these two samples are consistent with each other, but a slight different.
This may be caused by the variation of period, which will be discussed in the next section. Taking MJD 43840 as the phase 0, we show the folded phase histogram in Figure \ref{fig:foldedphase}.
The period is chosen as $395.86$ day, because the burst time of sample B is closer to phase 0.
In this figure, the blue histogram is the distribution of sample A and the red line is the kernel density estimation of sample B.
The distribution of these two samples both show a peak around phase 0.58. But in the case of sample A, the distribution
has other peaks, which are about 0.12, 0.77, and 0.87. While in sample B, the bursts are concentrated on phase 0.5-0.6. The number of
burst located in other phases is very small. The peak of phase distribution of these two different samples is similar, which enhances the reliable of the derived period.
Besides, the phase distribution of SGR 1806-20 is different from those of FRB 180916 and FRB 121102. The phase distributions of these two
FRBs are concentrated on a small region, while SGR 1806-20 has multiple peaks and spreads on the whole phase. We also show the burst
time and phase in Figure \ref{fig:mjd}. The different colored points denote the bursts observed by different telescopes.
The gray regions represent the main peak and the latter two peaks of the phases.
Most of the points are located in the gray region. Due to the existence
of the first peak, there are some points outside the gray region.

Although this period exists in these two samples, the significance of this peak is not strong enough. There are multiple peaks
	around 50-150 days in the Lomb-Scargle periodogram of sample A. Even considering the effects of the observation window, there are
	still several peaks that cannot be explained, such as the peaks about 76 days and 131 days. The reduce $ \chi^2 $ of epoch-folding
	also show several peaks about 149 days and 199 days. These peaks are difficult to understand. The results of sample B are much worse.
	The reduce $ \chi^2 $ and the Lomb-Scargle periodogram both have other significant peaks, and these peaks cannot be explained by the observation
	window.  Sample B only contains 208 bursts, which is much smaller than the bursts in sample A. The observation windows of sample B are not
	determined very well. Besides, the burst time of sample B spans a large range, from 1979 to 2005. The period has undergone evolution during
	this long epoch. All of these factors can have impacts on the results. In our results, 398 day is the most possible period of SGR 1806-20, but
	it is not significant enough.

\section{Discussions}\label{sec:diss}


The association between FRB 200428 and SGR 1935+2154 supports the conjecture that FRBs origin from magnetars
and FRBs are accompanied with X-ray bursts. Therefore, the periods of FRB and SGR
may be correlated.

Some theoretical models have been proposed to explain the periodic activity of FRBs. For example, the binary model has been proposed to explain the periods of FRB 180916 and FRB 121102 \citep{Ioka2020ApJ...893L..26I,Lyutikov2020ApJ...893L..39L}. Although this model can give a reasonable explanation of the period of FRBs, it is difficult to apply this model to SGRs.
There is no evidence to support the existence of a companion for SGR 1806-20. It is difficult to observe it due to the large distance. More importantly,
unlike the radio emission, the X-ray bursts would not be absorbed by stellar winds. Thus, the periodic activity of SGR 1806-20 can not be caused by the orbital motion.

Another promising model of FRBs period is the free precession of magnetars
\citep{Levin2020ApJ...895L..30L,Zanazzi2020ApJ...892L..15Z}. The free precession
originates from the non-sphericity of magnetars, which may be caused by the strong
internal magnetic field or the misaligned between the principal axis of the elastic crust
and the angular velocity \citep{Zanazzi2020ApJ...892L..15Z}. Although the superfluid vortices
insides the magnetar can suppress the free precession \citep{Shaham1977ApJ...214..251S}, \cite{Levin2020ApJ...895L..30L}
proposed that hyperactive magnetars are likely hot enough to quell superfluid vortices. Besides, the force precession
model also is discussed by some works \citep{Sobyanin2020MNRAS.497.1001S,Tong2020arXiv200210265T,Yang2020ApJ...893L..31Y}.
The torque can come from electromagnetic field of magnetars \citep{Sobyanin2020MNRAS.497.1001S}, the companion \citep{Yang2020ApJ...893L..31Y},
or the fallback disk \citep{Tong2020arXiv200210265T}. This torque can enhance the precession and lead to a large period.

In order to explain the periodic activity of FRBs, the free precession model requires that the radio bursts tend to occur in a specific location.
It is believed that X-ray bursts of SGRs are generated by starquakes of magnetars \citep{Thompson1995MNRAS.275..255T}.
But the trigger mechanism of these bursts is a mystery. Whether burst emissions locate in
a small region or a large area is unclear at present. Some models suggest that the bursts tend to
occur in a small region \citep{Gourgouliatos2015MNRAS.453L..93G,Lander2015MNRAS.449.2047L}. In this case,
the period of SGR 1806-20 can be explained by the free precession. The
precession period of magnetar is given by \citep{Levin2020ApJ...895L..30L}
\begin{equation}
    \label{eq:precession}
    P_{\mathrm{pr}} \simeq 396 \left(\frac{k}{0.01}\right)^{-1}\left(\frac{B_{\text {int}}}{7\times 10^{15} \mathrm{G}}\right)^{-2}\frac{B_{\text {dip}}}{2 \times 10^{15} \mathrm{G}}\left(\frac{t}{240~ \mathrm{yr}}\right)^{1 / 2} \text {days},
\end{equation}
where $k$ is numerical coefficient, $B_{\rm int}$ is the internal magnetic field of magnetar, $B_{\rm dip}$ is the surface dipole
magnetic field, and $t$ is the age of magnetar. If the magnetic field is fully coherent and purely toroidal, $k$ would approach 1. The
value of $k$ will be reduced if the field is tangled. The surface dipole magnetic field of SGR 1806-20 is $ 2 \times 10^{15} $ G and the age
is about 240 yr \citep{Olausen2014ApJS..212....6O}. If the internal magnetic field is about $ 7\times 10^{15} $ G and $k=0.01$, the precession period
is about 396 days, which is very close to the burst period of SGR 1806-20.

The precession model can also explain the wide span of bursts
in the phase space. First, these bursts are tend to occur in a special region, but also can occur in other positions. This will affect the period determination. Therefore, the bursts can span a wide range in the phase space. Second, from equation (\ref{eq:precession}), we can see that the precession period depends on the strength of magnetic field and age of magnetars. Therefore, the period can evolve. Because of the long observational time of SGR 1806-20, from 1979 to 2011, and the young age of SGR 1806-20, the period could evolve significantly, which causes some bursts outside the grey region in Figure \ref{fig:mjd}. The multiple peaks in Figure
\ref{fig:sampleA} and Figure \ref{fig:sampleB} may also be caused by the evolution of period.
 Considering the association between FRBs and X-ray bursts, if the precession explanation is correct, we predict that the periods of FRBs evolve with time. Some works suggested the ages of central magnetars of FRB 121102 and FRB 180916 are young \citep{Metzger2017,Cao2017,Marcote2020,Wu2020,Zhao2020}. Future long-term monitoring is required to test this prediction.



\section{Conclusion} \label{sec:conclusion}
The period behavior of FRBs is still a mystery. Given the connection between FRBs and X-ray bursts of SGRs, we investigate the
period behavior of SGR 1806-20. Three methods are used to derive the period, including the epoch folding method,
the Lomb-Scargle periodogram and the QMIEU periodogram. To alleviate the effect of unevenly sample, we divide the observation
into two samples. The sample A contains the observation of RXTE and the sample B includes the observation of
ICE, HETE-2 and Konus. We find the period $ 398.20 \pm 25.26 $ days for all the cases. The phase distribution
is shown in Figure \ref{fig:foldedphase}. The blue histogram is the distribution of sample A and the red line
is the kernel density estimate of sample B. The phase distribution is consistent with each other at the main
peak $\psi \simeq 0.58$.  There are other peaks in sample A, but these peaks are invisible in sample B.
Although the peak about 398 days is visible in both sample A and sample B, the existence of other peaks suggests
that this period is not significant enough. This period may be caused by observational window, not the internal period of SGR 1806-20.

We discuss possible physical mechanisms for the periodic behavior. If the triggers of bursts are tended to be localized in a small region,
the precession model can explain the periodic behavior. Considering the association between FRBs and bursts of SGRs, the physical mechanism
of periodic behavior may be same. The free precession model also predicts that the period evolves with time, which can be tested with
 long-term monitoring. The unstable period can also explain the multiple peaks in the periodogram and the bursts outside the main phase peak.

The association between SGR and FRB periods may be complex. From observations, 29 bursts of SGR 1935+2154 were not associated with FRBs \citep{Lin2020}. One most possible reason is that the FRB emission is much more beamed than SGR burst \citep{Zhang2020a}. Radio bursts from SGR J1935+2154 discovered by FAST \citep{2020ATel13699....1Z} and the BSA LPI radio telescope \citep{Alexander2020} may be due to beaming. If this situation is common in SGRs, the periods of  FRB activity and SGR activity are not the same. If the duty cycle of SGR bursts is large, there may be no relevant FRB period. On the other hand, several radio bursts were observed
overlapping with X-ray monitoring, without an associated
X-ray burst detection \citep{2020ATel13699....1Z,Kirsten2020}. Given the energy ratio between FRB
200428 and the XRB from SGR 1935+2154, the flux of X-ray burst is too low for current X-ray telescopes.

\section*{acknowledgements}
We thank the anonymous referee for constructive comments. We thank V. D. Pal'shin, for helpful discussion
on the observation window of Konus-Wind. We thank Z. Prieskorn and P. Kaaret for their kindness to share the burst catalog of SGR 1806-20. This work is supported by the National Natural Science Foundation of China (grant U1831207).

\bibliography{ms}{}

\begin{thebibliography}{}
\expandafter\ifx\csname natexlab\endcsname\relax\def\natexlab#1{#1}\fi
\providecommand{\url}[1]{\href{#1}{#1}}
\providecommand{\dodoi}[1]{doi:~\href{http://doi.org/#1}{\nolinkurl{#1}}}
\providecommand{\doeprint}[1]{\href{http://ascl.net/#1}{\nolinkurl{http://ascl.net/#1}}}
\providecommand{\doarXiv}[1]{\href{https://arxiv.org/abs/#1}{\nolinkurl{https://arxiv.org/abs/#1}}}

\bibitem[{{Alexander} \& {Fedorova}(2020)}]{Alexander2020}
{Alexander}, R., \& {Fedorova}, V. 2020, The Astronomer's Telegram, 14186, 1

\bibitem[{{Aptekar} {et~al.}(2001){Aptekar}, {Frederiks}, {Golenetskii},
  {Il'inskii}, {Mazets}, {Pal'shin}, {Butterworth}, \&
  {Cline}}]{Aptekar2001ApJS..137..227A}
{Aptekar}, R.~L., {Frederiks}, D.~D., {Golenetskii}, S.~V., {et~al.} 2001,
  \apjs, 137, 227, \dodoi{10.1086/322530}

\bibitem[{{Beloborodov}(2017)}]{Beloborodov2017ApJ...843L..26B}
{Beloborodov}, A.~M. 2017, \apjl, 843, L26, \dodoi{10.3847/2041-8213/aa78f3}

\bibitem[{{Bochenek} {et~al.}(2020){Bochenek}, {Ravi}, {Belov}, {Hallinan},
  {Kocz}, {Kulkarni}, \& {McKenna}}]{Bochenek2020arXiv200510828B}
{Bochenek}, C.~D., {Ravi}, V., {Belov}, K.~V., {et~al.} 2020, arXiv e-prints,
  arXiv:2005.10828.
\newblock \doarXiv{2005.10828}

\bibitem[{{Cao} {et~al.}(2017){Cao}, {Yu}, \& {Dai}}]{Cao2017}
{Cao}, X.-F., {Yu}, Y.-W., \& {Dai}, Z.-G. 2017, \apjl, 839, L20,
  \dodoi{10.3847/2041-8213/aa6af2}

\bibitem[{{Cheng} {et~al.}(2020){Cheng}, {Zhang}, \& {Wang}}]{Cheng2020}
{Cheng}, Y., {Zhang}, G.~Q., \& {Wang}, F.~Y. 2020, \mnras, 491, 1498,
  \dodoi{10.1093/mnras/stz3085}

\bibitem[{{CHIME/FRB Collaboration} {et~al.}(2020){CHIME/FRB Collaboration},
  {Amiri}, {Andersen}, {Band ura}, {Bhardwaj}, {Boyle}, {Brar}, {Chawla},
  {Chen}, {Cliche}, {Cubranic}, {Deng}, {Denman}, {Dobbs}, {Dong}, {Fand ino},
  {Fonseca}, {Gaensler}, {Giri}, {Good}, {Halpern}, {Hessels}, {Hill},
  {H{\"o}fer}, {Josephy}, {Kania}, {Karuppusamy}, {Kaspi}, {Keimpema},
  {Kirsten}, {Landecker}, {Lang}, {Leung}, {Li}, {Lin}, {Marcote}, {Masui},
  {McKinven}, {Mena-Parra}, {Merryfield}, {Michilli}, {Milutinovic},
  {Mirhosseini}, {Naidu}, {Newburgh}, {Ng}, {Nimmo}, {Paragi}, {Patel}, {Pen},
  {Pinsonneault-Marotte}, {Pleunis}, {Rafiei-Ravandi}, {Rahman}, {Ransom},
  {Renard}, {Sanghavi}, {Scholz}, {Shaw}, {Shin}, {Siegel}, {Singh}, {Smegal},
  {Smith}, {Stairs}, {Tendulkar}, {Tretyakov}, {Vanderlinde}, {Wang}, {Wang},
  {Wulf}, {Yadav}, \& {Zwaniga}}]{Chime/FrbCollaboration2020Natur.582..351C}
{CHIME/FRB Collaboration}, {Amiri}, M., {Andersen}, B.~C., {et~al.} 2020, \nat,
  582, 351, \dodoi{10.1038/s41586-020-2398-2}

\bibitem[{{Cordes} \& {Chatterjee}(2019)}]{Cordes2019ARA&A..57..417C}
{Cordes}, J.~M., \& {Chatterjee}, S. 2019, \araa, 57, 417,
  \dodoi{10.1146/annurev-astro-091918-104501}

\bibitem[{{Cruces} {et~al.}(2020){Cruces}, {Spitler}, {Scholz}, {Lynch},
  {Seymour}, {Hessels}, {Gouiff{\'e}s}, {Hilmarsson}, {Kramer}, \&
  {Munjal}}]{Cruces2020}
{Cruces}, M., {Spitler}, L.~G., {Scholz}, P., {et~al.} 2020, \mnras, 500, 448,
  \dodoi{10.1093/mnras/staa3223}

\bibitem[{{Dai} {et~al.}(2016){Dai}, {Wang}, {Wu}, \&
  {Huang}}]{Dai2016ApJ...829...27D}
{Dai}, Z.~G., {Wang}, J.~S., {Wu}, X.~F., \& {Huang}, Y.~F. 2016, \apj, 829,
  27, \dodoi{10.3847/0004-637X/829/1/27}

\bibitem[{{Dai} \& {Zhong}(2020)}]{Dai2020}
{Dai}, Z.~G., \& {Zhong}, S.~Q. 2020, \apjl, 895, L1,
  \dodoi{10.3847/2041-8213/ab8f2d}

\bibitem[{{Geng} \& {Huang}(2015)}]{Geng2015ApJ...809...24G}
{Geng}, J.~J., \& {Huang}, Y.~F. 2015, \apj, 809, 24,
  \dodoi{10.1088/0004-637X/809/1/24}

\bibitem[{{Gourgouliatos} {et~al.}(2015){Gourgouliatos}, {Kondi{\'c}},
  {Lyutikov}, \& {Hollerbach}}]{Gourgouliatos2015MNRAS.453L..93G}
{Gourgouliatos}, K.~N., {Kondi{\'c}}, T., {Lyutikov}, M., \& {Hollerbach}, R.
  2015, \mnras, 453, L93, \dodoi{10.1093/mnrasl/slv106}

\bibitem[{{Grossan}(2020)}]{Grossan2020arXiv200616480G}
{Grossan}, B. 2020, arXiv e-prints, arXiv:2006.16480.
\newblock \doarXiv{2006.16480}

\bibitem[{{Gu} {et~al.}(2020){Gu}, {Yi}, \& {Liu}}]{Gu2020}
{Gu}, W.-M., {Yi}, T., \& {Liu}, T. 2020, \mnras, 497, 1543,
  \dodoi{10.1093/mnras/staa1914}

\bibitem[{{Ioka} \& {Zhang}(2020)}]{Ioka2020ApJ...893L..26I}
{Ioka}, K., \& {Zhang}, B. 2020, \apjl, 893, L26,
  \dodoi{10.3847/2041-8213/ab83fb}

\bibitem[{{Israel} {et~al.}(2016){Israel}, {Esposito}, {Rea}, {Coti Zelati},
  {Tiengo}, {Campana}, {Mereghetti}, {Rodriguez Castillo}, {G{\"o}tz},
  {Burgay}, {Possenti}, {Zane}, {Turolla}, {Perna}, {Cannizzaro}, \&
  {Pons}}]{Israel2016MNRAS.457.3448I}
{Israel}, G.~L., {Esposito}, P., {Rea}, N., {et~al.} 2016, \mnras, 457, 3448,
  \dodoi{10.1093/mnras/stw008}

\bibitem[{{Kaspi} \& {Beloborodov}(2017)}]{Kaspi2017}
{Kaspi}, V.~M., \& {Beloborodov}, A.~M. 2017, \araa, 55, 261,
  \dodoi{10.1146/annurev-astro-081915-023329}

\bibitem[{{Katz}(2016)}]{Katz2016}
{Katz}, J.~I. 2016, \apj, 826, 226, \dodoi{10.3847/0004-637X/826/2/226}

\bibitem[{{Kirsten} {et~al.}(2020){Kirsten}, {Snelders}, {Jenkins}, {Nimmo},
  {van den Eijnden}, {Hessels}, {Gawro{\'n}ski}, \& {Yang}}]{Kirsten2020}
{Kirsten}, F., {Snelders}, M.~P., {Jenkins}, M., {et~al.} 2020, Nature
  Astronomy, \dodoi{10.1038/s41550-020-01246-3}

\bibitem[{{Kouveliotou} {et~al.}(1998){Kouveliotou}, {Dieters}, {Strohmayer},
  {van Paradijs}, {Fishman}, {Meegan}, {Hurley}, {Kommers}, {Smith}, {Frail},
  \& {Murakami}}]{Kouveliotou1998Natur.393..235K}
{Kouveliotou}, C., {Dieters}, S., {Strohmayer}, T., {et~al.} 1998, \nat, 393,
  235, \dodoi{10.1038/30410}

\bibitem[{{Kulkarni} {et~al.}(2014){Kulkarni}, {Ofek}, {Neill}, {Zheng}, \&
  {Juric}}]{Kulkarni2014}
{Kulkarni}, S.~R., {Ofek}, E.~O., {Neill}, J.~D., {Zheng}, Z., \& {Juric}, M.
  2014, \apj, 797, 70, \dodoi{10.1088/0004-637X/797/1/70}

\bibitem[{{Lander} {et~al.}(2015){Lander}, {Andersson}, {Antonopoulou}, \&
  {Watts}}]{Lander2015MNRAS.449.2047L}
{Lander}, S.~K., {Andersson}, N., {Antonopoulou}, D., \& {Watts}, A.~L. 2015,
  \mnras, 449, 2047, \dodoi{10.1093/mnras/stv432}

\bibitem[{{Laros} {et~al.}(1987){Laros}, {Fenimore}, {Klebesadel}, {Atteia},
  {Boer}, {Hurley}, {Niel}, {Vedrenne}, {Kane}, {Kouveliotou}, {Cline},
  {Dennis}, {Desai}, {Orwig}, {Kuznetsov}, {Sunyaev}, \&
  {Terekhov}}]{Laros1987ApJ...320L.111L}
{Laros}, J.~G., {Fenimore}, E.~E., {Klebesadel}, R.~W., {et~al.} 1987, \apjl,
  320, L111, \dodoi{10.1086/184985}

\bibitem[{{Levin} {et~al.}(2020){Levin}, {Beloborodov}, \&
  {Bransgrove}}]{Levin2020ApJ...895L..30L}
{Levin}, Y., {Beloborodov}, A.~M., \& {Bransgrove}, A. 2020, \apjl, 895, L30,
  \dodoi{10.3847/2041-8213/ab8c4c}

\bibitem[{{Li} {et~al.}(2020){Li}, {Lin}, {Xiong}, {Ge}, {Li}, {Li}, {Lu},
  {Zhang}, {Tuo}, {Nang}, {Zhang}, {Xiao}, {Chen}, {Song}, {Xu}, {Liu}, {Jia},
  {Cao}, {Zhang}, {Qu}, {Liao}, {Zhao}, {Tan}, {Nie}, {Zhao}, {Zheng}, {Zheng},
  {Luo}, {Cai}, {Li}, {Xue}, {Bu}, {Chang}, {Chen}, {Chen}, {Chen}, {Chen},
  {Chen}, {Cui}, {Cui}, {Deng}, {Dong}, {Du}, {Fu}, {Gao}, {Gao}, {Gao}, {Gu},
  {Guan}, {Guo}, {Han}, {Huang}, {Huo}, {Jiang}, {Jiang}, {Jin}, {Jin}, {Kong},
  {Li}, {Li}, {Li}, {Li}, {Li}, {Li}, {Li}, {Liang}, {Liu}, {Liu}, {Liu},
  {Liu}, {Liu}, {Lu}, {Lu}, {Luo}, {Ma}, {Meng}, {Ou}, {Sai}, {Shang}, {Song},
  {Sun}, {Tao}, {Wang}, {Wang}, {Wang}, {Wang}, {Wang}, {Wen}, {Wu}, {Wu},
  {Wu}, {Xiao}, {Yang}, {Yang}, {Yang}, {Yang}, {Yi}, {Yin}, {You}, {Zhang},
  {Zhang}, {Zhang}, {Zhang}, {Zhang}, {Zhang}, {Zhang}, {Zhang}, {Zhang},
  {Zhang}, {Zhang}, {Zhang}, {Zhang}, {Zhang}, {Zhang}, {Zhang}, {Zhou},
  {Zhou}, {Zhu}, {Zhu}, \& {Zhuang}}]{Li2020arXiv200511071L}
{Li}, C.~K., {Lin}, L., {Xiong}, S.~L., {et~al.} 2020, arXiv e-prints,
  arXiv:2005.11071.
\newblock \doarXiv{2005.11071}

\bibitem[{{Lin} {et~al.}(2020){Lin}, {Zhang}, {Wang}, {Gao}, {Guan}, {Han},
  {Jiang}, {Jiang}, {Lee}, {Li}, {Men}, {Miao}, {Niu}, {Niu}, {Sun}, {Wang},
  {Wang}, {Xu}, {Xu}, {Xu}, {Yang}, {Yang}, {Yu}, {Zhang}, {Zhang}, {Zhou},
  {Zhu}, {Castro-Tirado}, {Dai}, {Ge}, {Hu}, {Li}, {Li}, {Li}, {Liang}, {Jia},
  {Querel}, {Shao}, {Wang}, {Wang}, {Wu}, {Xiong}, {Xu}, {Yang}, {Zhang},
  {Zhang}, {Zheng}, \& {Zou}}]{Lin2020}
{Lin}, L., {Zhang}, C.~F., {Wang}, P., {et~al.} 2020, \nat, 587, 63,
  \dodoi{10.1038/s41586-020-2839-y}

\bibitem[{{Lomb}(1976)}]{Lomb1976Ap&SS..39..447L}
{Lomb}, N.~R. 1976, \apss, 39, 447, \dodoi{10.1007/BF00648343}

\bibitem[{{Lorimer} {et~al.}(2007){Lorimer}, {Bailes}, {McLaughlin},
  {Narkevic}, \& {Crawford}}]{Lorimer2007Sci...318..777L}
{Lorimer}, D.~R., {Bailes}, M., {McLaughlin}, M.~A., {Narkevic}, D.~J., \&
  {Crawford}, F. 2007, Science, 318, 777, \dodoi{10.1126/science.1147532}

\bibitem[{{Lyubarsky}(2014)}]{Lyubarsky2014MNRAS.442L...9L}
{Lyubarsky}, Y. 2014, \mnras, 442, L9, \dodoi{10.1093/mnrasl/slu046}

\bibitem[{{Lyutikov} {et~al.}(2020){Lyutikov}, {Barkov}, \&
  {Giannios}}]{Lyutikov2020ApJ...893L..39L}
{Lyutikov}, M., {Barkov}, M.~V., \& {Giannios}, D. 2020, \apjl, 893, L39,
  \dodoi{10.3847/2041-8213/ab87a4}

\bibitem[{{Marcote} {et~al.}(2020){Marcote}, {Nimmo}, {Hessels}, {Tendulkar},
  {Bassa}, {Paragi}, {Keimpema}, {Bhardwaj}, {Karuppusamy}, {Kaspi}, {Law},
  {Michilli}, {Aggarwal}, {Andersen}, {Archibald}, {Bandura}, {Bower}, {Boyle},
  {Brar}, {Burke-Spolaor}, {Butler}, {Cassanelli}, {Chawla}, {Demorest},
  {Dobbs}, {Fonseca}, {Giri}, {Good}, {Gourdji}, {Josephy}, {Kirichenko},
  {Kirsten}, {Landecker}, {Lang}, {Lazio}, {Li}, {Lin}, {Linford}, {Masui},
  {Mena-Parra}, {Naidu}, {Ng}, {Patel}, {Pen}, {Pleunis}, {Rafiei-Ravandi},
  {Rahman}, {Renard}, {Scholz}, {Siegel}, {Smith}, {Stairs}, {Vanderlinde}, \&
  {Zwaniga}}]{Marcote2020}
{Marcote}, B., {Nimmo}, K., {Hessels}, J.~W.~T., {et~al.} 2020, \nat, 577, 190,
  \dodoi{10.1038/s41586-019-1866-z}

\bibitem[{{Mereghetti} {et~al.}(2020){Mereghetti}, {Savchenko}, {Ferrigno},
  {G{\"o}tz}, {Rigoselli}, {Tiengo}, {Bazzano}, {Bozzo}, {Coleiro},
  {Courvoisier}, {Doyle}, {Goldwurm}, {Hanlon}, {Jourdain}, {Kienlin},
  {Lutovinov}, {Martin-Carrillo}, {Molkov}, {Natalucci}, {Onori}, {Panessa},
  {Rodi}, {Rodriguez}, {S{\'a}nchez-Fern{\'a}ndez}, {Sunyaev}, \&
  {Ubertini}}]{Mereghetti2020ApJ...898L..29M}
{Mereghetti}, S., {Savchenko}, V., {Ferrigno}, C., {et~al.} 2020, \apjl, 898,
  L29, \dodoi{10.3847/2041-8213/aba2cf}

\bibitem[{{Metzger} {et~al.}(2017){Metzger}, {Berger}, \&
  {Margalit}}]{Metzger2017}
{Metzger}, B.~D., {Berger}, E., \& {Margalit}, B. 2017, \apj, 841, 14,
  \dodoi{10.3847/1538-4357/aa633d}

\bibitem[{{Murase} {et~al.}(2016){Murase}, {Kashiyama}, \&
  {M{\'e}sz{\'a}ros}}]{Murase2016}
{Murase}, K., {Kashiyama}, K., \& {M{\'e}sz{\'a}ros}, P. 2016, \mnras, 461,
  1498, \dodoi{10.1093/mnras/stw1328}

\bibitem[{{Nakagawa} {et~al.}(2007){Nakagawa}, {Yoshida}, {Hurley}, {Atteia},
  {Maetou}, {Tamagawa}, {Suzuki}, {Yamazaki}, {Tanaka}, {Kawai}, {Shirasaki},
  {Pelangeon}, {Matsuoka}, {Vanderspek}, {Crew}, {Villasenor}, {Sato},
  {Sugita}, {Kotoku}, {Arimoto}, {Pizzichini}, {Doty}, \&
  {Ricker}}]{Nakagawa2007PASJ...59..653N}
{Nakagawa}, Y.~E., {Yoshida}, A., {Hurley}, K., {et~al.} 2007, \pasj, 59, 653,
  \dodoi{10.1093/pasj/59.3.653}

\bibitem[{{Olausen} \& {Kaspi}(2014)}]{Olausen2014ApJS..212....6O}
{Olausen}, S.~A., \& {Kaspi}, V.~M. 2014, \apjs, 212, 6,
  \dodoi{10.1088/0067-0049/212/1/6}

\bibitem[{{Palmer} {et~al.}(2005){Palmer}, {Barthelmy}, {Gehrels}, {Kippen},
  {Cayton}, {Kouveliotou}, {Eichler}, {Wijers}, {Woods}, {Granot}, {Lyubarsky},
  {Ramirez-Ruiz}, {Barbier}, {Chester}, {Cummings}, {Fenimore}, {Finger},
  {Gaensler}, {Hullinger}, {Krimm}, {Markwardt}, {Nousek}, {Parsons}, {Patel},
  {Sakamoto}, {Sato}, {Suzuki}, \& {Tueller}}]{Palmer2005Natur.434.1107P}
{Palmer}, D.~M., {Barthelmy}, S., {Gehrels}, N., {et~al.} 2005, \nat, 434,
  1107, \dodoi{10.1038/nature03525}

\bibitem[{{Petroff} {et~al.}(2019){Petroff}, {Hessels}, \&
  {Lorimer}}]{Petroff2019A&ARv..27....4P}
{Petroff}, E., {Hessels}, J.~W.~T., \& {Lorimer}, D.~R. 2019, \aapr, 27, 4,
  \dodoi{10.1007/s00159-019-0116-6}

\bibitem[{{Platts} {et~al.}(2019){Platts}, {Weltman}, {Walters}, {Tendulkar},
  {Gordin}, \& {Kandhai}}]{Platts2019PhR...821....1P}
{Platts}, E., {Weltman}, A., {Walters}, A., {et~al.} 2019, \physrep, 821, 1,
  \dodoi{10.1016/j.physrep.2019.06.003}

\bibitem[{{Prieskorn} \& {Kaaret}(2012)}]{Prieskorn2012ApJ...755....1P}
{Prieskorn}, Z., \& {Kaaret}, P. 2012, \apj, 755, 1,
  \dodoi{10.1088/0004-637X/755/1/1}

\bibitem[{{Rajwade} {et~al.}(2020){Rajwade}, {Mickaliger}, {Stappers},
  {Morello}, {Agarwal}, {Bassa}, {Breton}, {Caleb}, {Karastergiou}, {Keane}, \&
  {Lorimer}}]{Rajwade2020MNRAS.495.3551R}
{Rajwade}, K.~M., {Mickaliger}, M.~B., {Stappers}, B.~W., {et~al.} 2020,
  \mnras, 495, 3551, \dodoi{10.1093/mnras/staa1237}

\bibitem[{{Scargle}(1982)}]{Scargle1982ApJ...263..835S}
{Scargle}, J.~D. 1982, \apj, 263, 835, \dodoi{10.1086/160554}

\bibitem[{{Shaham}(1977)}]{Shaham1977ApJ...214..251S}
{Shaham}, J. 1977, \apj, 214, 251, \dodoi{10.1086/155249}

\bibitem[{{Sob'yanin}(2020)}]{Sobyanin2020MNRAS.497.1001S}
{Sob'yanin}, D.~N. 2020, \mnras, 497, 1001, \dodoi{10.1093/mnras/staa1976}

\bibitem[{{Tendulkar} {et~al.}(2016){Tendulkar}, {Kaspi}, \&
  {Patel}}]{Tendulkar2016ApJ...827...59T}
{Tendulkar}, S.~P., {Kaspi}, V.~M., \& {Patel}, C. 2016, \apj, 827, 59,
  \dodoi{10.3847/0004-637X/827/1/59}

\bibitem[{{The CHIME/FRB Collaboration} {et~al.}(2020){The CHIME/FRB
  Collaboration}, {:}, {Andersen}, {Band ura}, {Bhardwaj}, {Bij}, {Boyce},
  {Boyle}, {Brar}, {Cassanelli}, {Chawla}, {Chen}, {Cliche}, {Cook},
  {Cubranic}, {Curtin}, {Denman}, {Dobbs}, {Dong}, {Fandino}, {Fonseca},
  {Gaensler}, {Giri}, {Good}, {Halpern}, {Hill}, {Hinshaw}, {H{\"o}fer},
  {Josephy}, {Kania}, {Kaspi}, {Landecker}, {Leung}, {Li}, {Lin}, {Masui},
  {Mckinven}, {Mena-Parra}, {Merryfield}, {Meyers}, {Michilli}, {Milutinovic},
  {Mirhosseini}, {M{\"u}nchmeyer}, {Naidu}, {Newburgh}, {Ng}, {Patel}, {Pen},
  {Pinsonneault-Marotte}, {Pleunis}, {Quine}, {Rafiei-Ravandi}, {Rahman},
  {Ransom}, {Renard}, {Sanghavi}, {Scholz}, {Shaw}, {Shin}, {Siegel}, {Singh},
  {Smegal}, {Smith}, {Stairs}, {Tan}, {Tendulkar}, {Tretyakov}, {Vanderlinde},
  {Wang}, {Wulf}, \& {Zwaniga}}]{TheCHIME/FRBCollaboration2020arXiv200510324T}
{The CHIME/FRB Collaboration}, {:}, {Andersen}, B.~C., {et~al.} 2020, arXiv
  e-prints, arXiv:2005.10324.
\newblock \doarXiv{2005.10324}

\bibitem[{{Thompson} \& {Duncan}(1995)}]{Thompson1995MNRAS.275..255T}
{Thompson}, C., \& {Duncan}, R.~C. 1995, \mnras, 275, 255,
  \dodoi{10.1093/mnras/275.2.255}

\bibitem[{{Tong} {et~al.}(2020){Tong}, {Wang}, \&
  {Wang}}]{Tong2020arXiv200210265T}
{Tong}, H., {Wang}, W., \& {Wang}, H.~G. 2020, arXiv e-prints,
  arXiv:2002.10265.
\newblock \doarXiv{2002.10265}

\bibitem[{{Ulmer} {et~al.}(1993){Ulmer}, {Fenimore}, {Epstein}, {Ho},
  {Klebesadel}, {Laros}, \& {Delgado}}]{Ulmer1993ApJ...418..395U}
{Ulmer}, A., {Fenimore}, E.~E., {Epstein}, R.~I., {et~al.} 1993, \apj, 418,
  395, \dodoi{10.1086/173399}

\bibitem[{{VanderPlas}(2018)}]{VanderPlas2018ApJS..236...16V}
{VanderPlas}, J.~T. 2018, \apjs, 236, 16, \dodoi{10.3847/1538-4365/aab766}

\bibitem[{{Wadiasingh} \& {Timokhin}(2019)}]{Wadiasingh2019}
{Wadiasingh}, Z., \& {Timokhin}, A. 2019, \apj, 879, 4,
  \dodoi{10.3847/1538-4357/ab2240}

\bibitem[{{Wang} {et~al.}(2020){Wang}, {Wang}, {Yang}, {Yu}, {Zuo}, \&
  {Dai}}]{Wang2020}
{Wang}, F.~Y., {Wang}, Y.~Y., {Yang}, Y.-P., {et~al.} 2020, \apj, 891, 72,
  \dodoi{10.3847/1538-4357/ab74d0}

\bibitem[{{Wang} \& {Yu}(2017)}]{Wang2017}
{Wang}, F.~Y., \& {Yu}, H. 2017, \jcap, 03, 023,
  \dodoi{10.1088/1475-7516/2017/03/023}

\bibitem[{{Woods} {et~al.}(2007){Woods}, {Kouveliotou}, {Finger},
  {G{\"o}{\v{g}}{\"u}{\textcommabelow s}}, {Wilson}, {Patel}, {Hurley}, \&
  {Swank}}]{Woods2007ApJ...654..470W}
{Woods}, P.~M., {Kouveliotou}, C., {Finger}, M.~H., {et~al.} 2007, \apj, 654,
  470, \dodoi{10.1086/507459}

\bibitem[{{Wu} {et~al.}(2020){Wu}, {Zhang}, {Wang}, \& {Dai}}]{Wu2020}
{Wu}, Q., {Zhang}, G.~Q., {Wang}, F.~Y., \& {Dai}, Z.~G. 2020, arXiv e-prints,
  arXiv:2008.05635.
\newblock \doarXiv{2008.05635}

\bibitem[{{Xiao} {et~al.}(2021){Xiao}, {Wang}, \& {Dai}}]{Xiao2021}
{Xiao}, D., {Wang}, F., \& {Dai}, Z. 2021, arXiv e-prints, arXiv:2101.04907.
\newblock \doarXiv{2101.04907}

\bibitem[{{Yang} \& {Zou}(2020)}]{Yang2020ApJ...893L..31Y}
{Yang}, H., \& {Zou}, Y.-C. 2020, \apjl, 893, L31,
  \dodoi{10.3847/2041-8213/ab800f}

\bibitem[{{Zanazzi} \& {Lai}(2020)}]{Zanazzi2020ApJ...892L..15Z}
{Zanazzi}, J.~J., \& {Lai}, D. 2020, \apjl, 892, L15,
  \dodoi{10.3847/2041-8213/ab7cdd}

\bibitem[{{Zhang}(2017)}]{Zhang2017ApJ...836L..32Z}
{Zhang}, B. 2017, \apjl, 836, L32, \dodoi{10.3847/2041-8213/aa5ded}

\bibitem[{{Zhang}(2018)}]{Zhang2018ApJ...854L..21Z}
---. 2018, \apjl, 854, L21, \dodoi{10.3847/2041-8213/aaadba}

\bibitem[{{Zhang}(2020{\natexlab{a}})}]{Zhang2020}
---. 2020{\natexlab{a}}, \nat, 587, 45, \dodoi{10.1038/s41586-020-2828-1}

\bibitem[{{Zhang}(2020{\natexlab{b}})}]{Zhang2020a}
---. 2020{\natexlab{b}}, arXiv e-prints, arXiv:2011.09921.
\newblock \doarXiv{2011.09921}

\bibitem[{{Zhang} {et~al.}(2020){Zhang}, {Jiang}, {Men}, {Wang}, {Xu}, {Xu},
  {Niu}, {Zhou}, {Guan}, {Han}, {Jiang}, {Lee}, {Li}, {Lin}, {Niu}, {Wang},
  {Wang}, {Xu}, {Yu}, {Zhang}, \& {Zhu}}]{2020ATel13699....1Z}
{Zhang}, C.~F., {Jiang}, J.~C., {Men}, Y.~P., {et~al.} 2020, The Astronomer's
  Telegram, 13699, 1

\bibitem[{{Zhao} {et~al.}(2020){Zhao}, {Zhang}, {Wang}, \& {Wang}}]{Zhao2020}
{Zhao}, Z.~Y., {Zhang}, G.~Q., {Wang}, Y.~Y., \& {Wang}, F.~Y. 2020, arXiv
  e-prints, arXiv:2010.10702.
\newblock \doarXiv{2010.10702}

\end{thebibliography}
\bibliographystyle{aasjournal}

\clearpage

\end{document}